\documentclass[sigconf, nonacm]{acmart}

\AtBeginDocument{%
  \providecommand\BibTeX{{%
    Bib\TeX}}}

\copyrightyear{2025} 
\acmYear{2025} 
\setcopyright{rightsretained} 
\acmConference[]{}{}{}
\acmBooktitle{}

\author{Kristian Kolthoff}
\authornotemark[1]
\orcid{0000-0003-4982-488X}

\affiliation{%
  \institution{Software and Systems Engineering}
  \institution{Clausthal University of Technology}
  \city{Clausthal-Zellerfeld}
  \country{Germany}
}
\email{kristian.kolthoff@tu-clausthal.de}

\author{Felix Kretzer}
\authornote{Both authors contributed equally to the paper.}

\orcid{0000-0001-8115-7592}

\affiliation{%
\institution{human-centered systems lab (h-lab)}
  \institution{Karlsruhe Institute of Technology}
  \city{Karlsruhe}
  \country{Germany}
}
\email{felix.kretzer@kit.edu}

\author{Simone Paolo Ponzetto}
\orcid{0000-0001-7484-2049}

\affiliation{%
  \institution{Data and Web Science Group University of Mannheim}
  \city{Mannheim}
  \country{Germany}
}
\email{ponzetto@uni-mannheim.de}

\author{Alexander Maedche}
\orcid{0000-0001-6546-4816}

\affiliation{%
  \institution{human-centered systems lab (h-lab)}
  \institution{Karlsruhe Institute of Technology}
  \city{Karlsruhe}
  \country{Germany}
}
\email{alexander.maedche@kit.edu}

\author{Christian Bartelt}
\orcid{0000-0003-0426-6714}

\affiliation{%
  \institution{Software and Systems Engineering}
  \institution{Clausthal University of Technology}
  \city{Clausthal-Zellerfeld}
  \country{Germany}
}
\email{christian.bartelt@tu-clausthal.de}


\usepackage{graphicx}
\usepackage{textcomp}
\usepackage{enumitem}
\usepackage{multirow}
\usepackage{colortbl}
\usepackage{array}
\usepackage{tcolorbox}
\usepackage{nicematrix}
\usepackage{tabularx}
\usepackage{balance}
\usepackage{colortbl}
\usepackage{booktabs}
\usepackage{url}
\usepackage{calc}
\usepackage{siunitx}
\usepackage{makecell}

\newcommand{\pval}[2]{%
  \ifnum#1<50 \textbf{\small (#2)} \else \small (#2) \fi
}

\definecolor{lightgray}{gray}{0.9}

\PassOptionsToPackage{normalem}{ulem}
\usepackage{ulem}

\providecolor{added}{rgb}{0,0,1}
\providecolor{deleted}{rgb}{1,0,0}

\usepackage{subcaption}
\usepackage{mwe}
\usepackage{hyperref}

\hypersetup{
    colorlinks=false,
    pdfborder={0 0 0}
}

\definecolor{specialblue}{RGB}{0,104,149}
\definecolor{specialgray}{RGB}{242,242,242}

\usepackage{gb4e}
\noautomath

\usepackage{tikz}
\usepackage[framemethod=tikz]{mdframed}
\usetikzlibrary{calc}

\newtcolorbox{mybox}{
  sharp corners,
  colback=specialgray,
  colframe=specialblue,
  boxrule=0pt,
    toprule=0pt,
  bottomrule=0pt,
  leftrule=3pt, 
  rightrule=3pt 
}

\usepackage{cite}
\usepackage{amsmath,amsfonts}
\usepackage{algorithmic}
\def\BibTeX{{\rm B\kern-.05em{\sc i\kern-.025em b}\kern-.08em
    T\kern-.1667em\lower.7ex\hbox{E}\kern-.125emX}}

\usepackage{eso-pic}

\begin{document}

\title{GUISpector: An MLLM Agent Framework for Automated Verification of Natural Language Requirements in GUI Prototypes}


\renewcommand{\shortauthors}{Kolthoff and Kretzer et al.}
\begin{abstract}
Graphical user interfaces (GUIs) are foundational to interactive systems and play a pivotal role in early requirements elicitation through prototyping. Ensuring that GUI implementations fulfill natural language (NL) requirements is essential for robust software engineering, especially as LLM-driven programming agents become increasingly integrated into development workflows. Existing GUI testing approaches, whether traditional or LLM-driven, often fall short in handling the complexity of modern interfaces, and typically lack actionable feedback and effective integration with automated development agents. In this paper, we introduce \textit{GUISpector}, a novel framework that leverages a multi-modal (M)LLM-based agent for the automated verification of NL requirements in GUI prototypes. 
First, \textit{GUISpector} adapts a MLLM agent to interpret and operationalize NL requirements, enabling to autonomously plan and execute verification trajectories across GUI applications. 
Second, \textit{GUISpector} systematically extracts detailed NL feedback from the agent’s verification process, providing developers with actionable insights that can be used to iteratively refine the GUI artifact or directly inform LLM-based code generation in a closed feedback loop. 
Third, we present an integrated tool that unifies these capabilities, offering practitioners an accessible interface for supervising verification runs, inspecting agent rationales and managing the end-to-end requirements verification process. 
We evaluated \textit{GUISpector} on a comprehensive set of 150 requirements based on 900 acceptance criteria annotations across diverse GUI applications, demonstrating effective detection of requirement satisfaction and violations and highlighting its potential for seamless integration of actionable feedback into automated LLM-driven development workflows. The video presentation of \textit{GUISpector} is available at: \url{https://youtu.be/JByYF6BNQeE}, showcasing its main capabilities.
\end{abstract}



\maketitle

\section{Introduction}

Graphical user interfaces (GUIs) are central to effective interaction with modern software systems and have become ubiquitous across domains. Beyond their role in end-user interaction, GUIs are frequently employed as prototypes during requirements elicitation \citep{pohl2010requirements}, providing stakeholders with concrete, interactive artifacts that facilitate clearer communication and more precise specification of requirements \citep{ravid2000method, beaudouin2002prototyping}. As with other critical software artifacts, rigorous GUI testing is essential to ensure both functional correctness and stakeholder alignment. However, GUIs often exhibit significant complexity and dynamic behavior, making comprehensive testing a challenging and resource-intensive task. Moreover, human test data is often ambiguous and existing approaches rely on often scattered tools \citep{kretzer_tesy}. Ensuring that GUIs correctly implement elicited requirements is particularly important in the prototyping phase, as discrepancies can lead to costly misunderstandings and rework \citep{beaudouin2002prototyping}. The recent surge in large language model (LLM)-driven GUI generation \citep{kolthoff2024zero, fiebig2025effective} has accelerated automated GUI development, making equally robust automated testing essential.

\begin{figure*}[!t]
  \centering
 \includegraphics[width=0.955\textwidth]{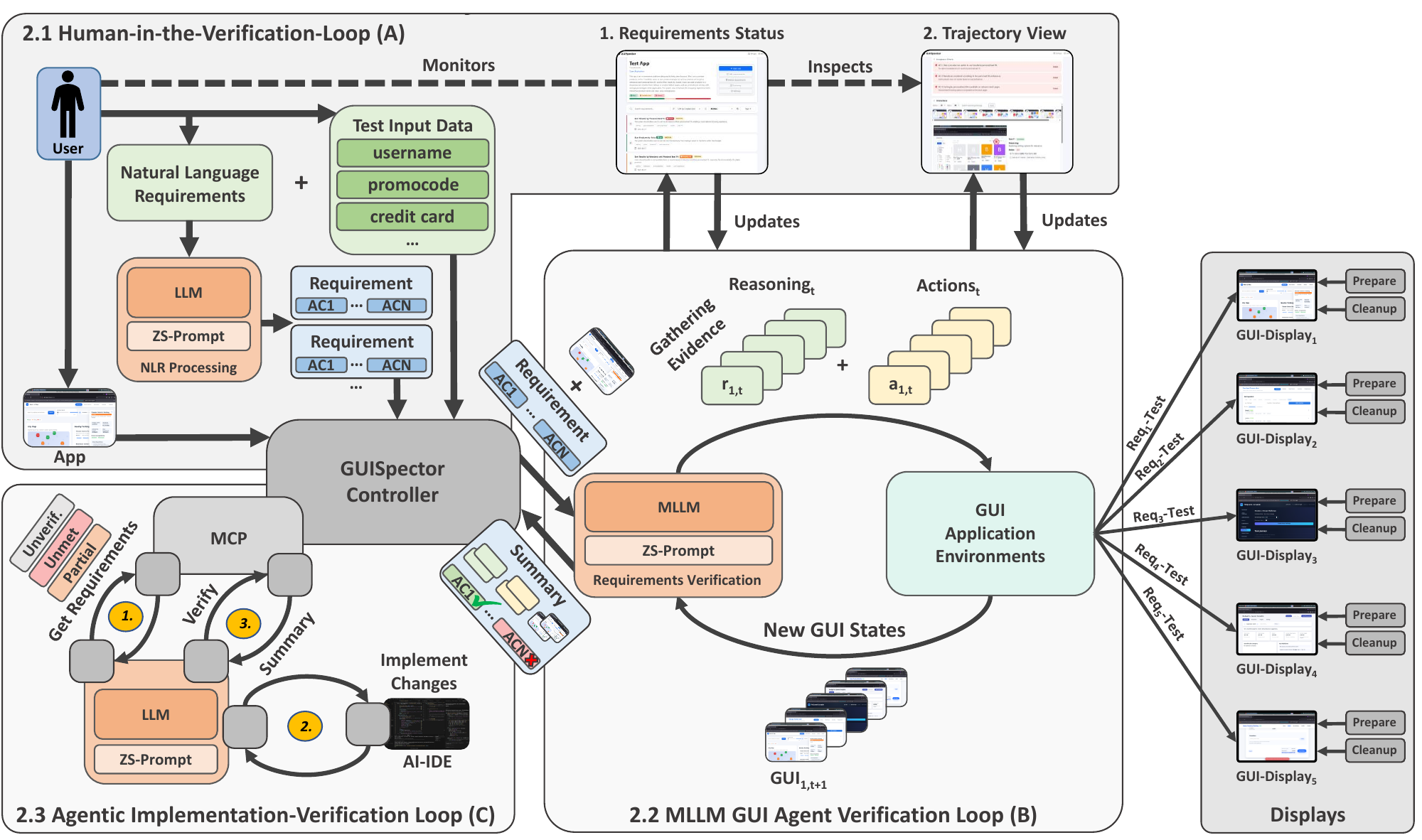}
  \caption{Overview of the \textit{GUISpector} architecture with Human-in-the-Verification-Loop, MLLM GUI agent verification loop with reasoning, actions and GUI state trajectories and agentic implementation-verification loop for fully autonomous mode}
  \Description{}
  \label{fig:overview}
\end{figure*}

To alleviate the burden of manual GUI testing, numerous automated approaches have been proposed, including automated test scripts \citep{xie2007designing}, random \citep{mao2016sapienz} and model-based exploration for improved coverage \citep{gu2019practical, zeng2016automated}, and scripted bug replay techniques \citep{gomez2013reran, feng2022gifdroid}. However, these methods often struggle with low coverage due to the complexity and dynamic nature of GUIs. Recent advances leverage LLMs to enable more sophisticated and semantically meaningful GUI testing. For example, LLMs have been used to predict user actions for state exploration \citep{liu2023chatting} and to generate tests from natural language descriptions \citep{feng2024enabling, feng2025agent}. Yet, these approaches typically rely on textual abstractions of GUI hierarchies, missing critical dynamic and visual aspects and are unable to address non-functional and fine-grained requirements.

In this paper, we address this gap by introducing \textit{GUISpector}, a multi-modal (M)LLM-based agent adapted for the automated verification of natural language requirements in GUI prototypes. \textit{GUISpector} leverages vision-language reasoning to interpret and operationalize NL requirements, autonomously explore GUI applications and extract actionable summaries from test trajectories. Crucially, our approach enables a closed feedback loop, where insights from GUI verification can be directly fed back into LLM-driven GUI generation, a capability that remains largely unexplored despite its demonstrated benefits in code generation workflows \citep{peng2025perfcodegen}. As LLM-driven GUI development becomes widely adopted \citep{kolthoff2024zero, fiebig2025effective}, \textit{GUISpector} provides an accessible, integrated tool that empowers practitioners to seamlessly incorporate requirements-driven verification and iterative improvement into automated GUI development. Code, datasets and prototype are available at our repository \citep{github}.
\section{Approach: GUISpector}

\begin{table*}[!t]
\footnotesize
\caption[Evaluation Results]{Evaluation results for requirements (Req) and acceptance criteria (AC) across five applications, with averages of steps, time and token consumption. Token counts are reported in thousands (k). Costs using \textit{OpenAI CUA}: \$3/M input, \$12/M output. Number of steps, time, input and output-token consumption, and costs are reported per requirement, all other metrics per app.}
\centering
\setlength\tabcolsep{3.1pt}
\renewcommand{\arraystretch}{1.25}
\begin{tabular}{S||ccc|ccc|ccc|ccc|ccc||cc|cc|cc|cc|cc}
\toprule
& \multicolumn{3}{c|}{\textbf{Met (Req)}} 
& \multicolumn{3}{c|}{\textbf{Unmet (Req)}} 
& \multicolumn{3}{c|}{\textbf{Partial (Req)}} 
& \multicolumn{3}{c|}{\textbf{Met (AC)}} 
& \multicolumn{3}{c||}{\textbf{Unmet (AC)}} 
& \multicolumn{2}{c|}{\textbf{\#Steps}} 
& \multicolumn{2}{c|}{\textbf{Time (s)}} 
& \multicolumn{2}{c|}{\textbf{\#In-Tok. (k)}} 
& \multicolumn{2}{c|}{\textbf{\#Out-Tok. (k)}} 
& \multicolumn{2}{c}{\textbf{Cost (\$)}} \\
\cmidrule(lr){2-4} \cmidrule(lr){5-7} \cmidrule(lr){8-10} \cmidrule(lr){11-13} \cmidrule(lr){14-16}
\cmidrule(lr){17-18} \cmidrule(lr){19-20} \cmidrule(lr){21-22} \cmidrule(lr){23-24} \cmidrule(l){25-26}
& \textbf{P} & \textbf{R} & \textbf{F1} 
& \textbf{P} & \textbf{R} & \textbf{F1} 
& \textbf{P} & \textbf{R} & \textbf{F1} 
& \textbf{P} & \textbf{R} & \textbf{F1} 
& \textbf{P} & \textbf{R} & \textbf{F1} 
& \textbf{Av.} & \textbf{SD}
& \textbf{Av.} & \textbf{SD}
& \textbf{Av.} & \textbf{SD}
& \textbf{Av.} & \textbf{SD}
& \textbf{Av.} & \textbf{SD} \\
\midrule
\textbf{App-1}  & .789 & .938 & .857 & 1.00 & .714 & .833 & .500 & .429 & .462 & .908 & .967 & .937 & .920 & .793 & .852 & 25.1 & 17.4 & 336.0 & 291.2 & 229.8 & 165.5 & 2.394 & 1.323 & .689 & .029 \\
\rowcolor{lightgray}
\textbf{App-2}  & 1.00 & .944 & .971 & 1.00 & 1.00 & 1.00 & .833 & 1.00 & .909 & .968 & .984 & .976 & .958 & .920 & .939 & 21.1 & 14.9 & 322.3 & 316.0 & 189.7 & 127.8 & 1.984 & 0.817 & .569 & .024 \\
\textbf{App-3}  & .933 & .824 & .875 & .833 & .833 & .833 & .625 & .833 & .714 & .931 & .900 & .915 & .793 & .852 & .821 & 19.5 & 16.6 & 223.1 & 240.5 & 174.6 & 149.2 & 2.002 & 1.020 & .524 & .024 \\
\rowcolor{lightgray}
\textbf{App-4}  & 1.00 & .824 & .903 & .750 & .857 & .800 & .500 & .667 & .571 & .946 & .898 & .922 & .824 & .903 & .862 & 22.3 & 12.4 & 293.4 & 182.3 & 202.7 & 107.7 & 2.115 & 0.822 & .608 & .025 \\
\textbf{App-5}  & .850 & 1.00 & .919 & 1.00 & .857 & .923 & .667 & .400 & .500 & .909 & 1.00 & .952 & 1.00 & .778 & .875 & 34.2 & 19.8 & 412.9 & 266.7 & 311.9 & 188.0 & 2.850 & 1.399 & .936 & .034 \\
\midrule
\rowcolor{lightgray}
\textbf{Avg.}  & .915 & .906 & .905 & .917 & .852 & .878 & .625 & .666 & .631 & .933 & .950 & .940 & .899 & .849 & .870 & 24.4 & 16.2 & 317.6 & 259.3 & 221.7 & 147.6 & 2.269 & 1.076 & .665 & .027 \\
\textbf{SD}  & .093 & .079 & .044 & .118 & .102 & .082 & .138 & .258 & .183 & .026 & .048 & .025 & .088 & .064 & .043 & 5.8 & 2.8 & 68.8 & 51.4 & 54.3 & 31.4 & 0.364 & 0.274 & .163 & .004 \\
\bottomrule
\end{tabular}
\label{tab:results_nlr_pm_steps}
\end{table*}

\textit{GUISpector} comprises three main components: \textit{(A)} a human-in-the-verification-loop, enabling users to specify the target application, natural language requirements, input data, and monitoring the verification process, \textit{(B)} the core MLLM GUI agent verification loop, iteratively generating reasoning for planning and evaluation, selecting actions, and recording GUI screenshot trajectories, and \textit{(C)} an agentic implementation-verification loop that enables LLM-based programming agents to seamlessly interact with \textit{GUISpector}, for fully automated workflows. The architecture is shown in Fig.~\ref{fig:overview}.

\subsection{Human-in-the-Verification-Loop}

In \textit{GUISpector}, users begin by specifying the application for verification, providing unstructured NL requirements and supplying relevant test input data. Requirements are automatically processed using a zero-shot prompted LLM, which extracts a structured representation of each requirement. This acts as a generic NL interface, capturing key elements such as the description and particularly individual acceptance criteria, with the option to infer missing details from context. \textit{GUISpector} offers an intuitive user interface that presents an overview of all requirements along with their current fulfillment status and allows users to inspect the verification trajectories executed by the MLLM GUI agent. Each agent interaction can be examined in detail, including the MLLM reasoning and a breakdown of acceptance criteria fulfillment or violation, supported by evidence collected during verification runs. Overall, \textit{GUISpector} supports efficient, parallelized NL requirements-driven GUI testing.

\subsection{MLLM GUI Agent Verification Loop}

The core component of \textit{GUISpector} is the MLLM agent verification loop, which adapts a pretrained multi-modal LLM optimized for computer use by employing a zero-shot prompt that instructs the agent to autonomously verify whether each requirement is implemented in the application, starting from a given URL and interacting with the GUI to gather evidence. The prompt enforces fully autonomous operation, minimal actions, use of user-provided input test data and requires the agent to evaluate each acceptance criterion individually, justifying its decision with concrete evidence and outputting a structured JSON summary. For each verification run, the agent receives the requirement, acceptance criteria and test data along with a GUI screenshot representing the current application state. The agent then reasons about the most plausible next actions to verify acceptance criteria, predicts the next action (e.g., \textit{click}, \textit{double click}, \textit{scroll}, \textit{type}, with parameters such as \textit{x,y} coordinates), and executes it in the GUI environment, which updates the GUI state. Subsequently, a new screenshot is captured and returned to the model, closing the loop. This iterative process produces a trajectory of GUI states, reasoning steps, and actions (\textit{GUI}$_1$, r$_1$, a$_1$, \ldots, \textit{GUI}$_n$, r$_n$, a$_n$), culminating in a final output JSON containing a detailed summary, explanations and evidence for each acceptance criterion. These summaries (particularly important for violated requirements) can be accessed by users and fed back into LLM-driven programming agents to support iterative GUI development and improvement. Before and after each verification run, the GUI environments are cleaned up to avoid contaminating states.

\subsection{Agentic Implementation-Verification Loop}

Beyond manual user-driven requirements verification, \textit{GUISpector} also supports a fully automatic mode that integrates LLM-based programming agents with the proposed GUI verification approach closing the development loop. With the emergence and recent popularity of LLM-powered IDEs such as \textit{Cursor} \citep{cursor} and \textit{Copilot} \citep{githubcopilot}, developers can easily integrate their large-scale code bases into LLMs, enabling to rapidly implement changes with LLMs and potentially increase productivity. To facilitate seamless integration, we implemented a \textit{Model Context Protocol (MCP)} \citep{hou2025model} server, allowing external agents to directly interact with \textit{GUISpector}. LLM-driven programming agents can retrieve requirements, including acceptance criteria and their current verification states for different verification setups previously created by the user and selectively process unverified or violated requirements. Agents can then invoke verification runs through \textit{GUISpector}, receiving actionable explanations for any \textit{unmet} or \textit{partially met} requirements. This enables iterative improvement: after addressing identified issues, the agent re-runs verification until the requirement is \textit{met}, then proceeds to the next one. We employ a zero-shot prompt to instruct programming agents to follow this fully automated implementation-verification loop, representing a novel approach that leverages actionable feedback from automated MLLM GUI agents to drive iterative, requirements-driven development and enhance the capabilities of LLM agents.

\subsection{Implementation}

\textit{GUISpector} is built as a \textit{Django} web application with an \textit{HTML/CSS} and \textit{JavaScript} frontend and uses \textit{MySQL} for data storage. Background tasks, such as executing parallel MLLM agent verifications, are managed with \textit{Celery} across multiple worker nodes, coordinated by \textit{Redis} for efficient and scalable processing. To support simultaneous verification runs, we implemented a display resource manager in \textit{Redis} that dynamically allocates and tracks \textit{Xfce} desktop environments running on \textit{Xvfb} virtual displays. \textit{xdotool} is used to simulate user interactions within GUI applications. For MLLM-based verification, we currently employ the \textit{OpenAI CUA} \citep{cua2025} model, while for other LLM tasks, the framework offers support for \textit{OpenAI GPT} \citep{openai2023gpt4}, \textit{Google Gemini} \citep{team2023gemini}, and \textit{Anthropic Claude Sonnet} \citep{claude4}, allowing flexible model selection. Other MLLM computer-use agents can seamlessly be integrated into the proposed \textit{GUISpector} framework.
\section{Experimental Evaluation}

\vspace{0.2cm}\noindent
\textit{GUI-Requirements Evaluation Dataset.} 
To evaluate the ability of the proposed approach to recognize requirements fulfillment, we first created a new dataset, due to the lack of available datasets of GUIs paired with requirements. Using \textit{Codex} \citep{codex2025}, we generated 30 functional requirements with three acceptance criteria (ACs) each for five domain-diverse applications (\textit{Park-and-Pay, Budget Tracker, Recipe Generator, Fitness Quests, Cleaning Booking}). In a second step, \textit{Codex} was instructed to create the corresponding apps (single HTML files with local state, no external CDNs/APIs). All prompts, requirements, ACs and apps are available in our repository \citep{github}.

We then evaluated whether the generated apps conformed to the specifications. For each app, the target distribution was 18  \textit{met}, six \textit{unmet} and six \textit{partially met} requirements (30 per app). Two authors independently labeled 450 acceptance criteria (ACs) as \textit{met} or \textit{unmet}. Annotations were independent but not equally blinded: one annotator had not seen the detailed specification; the other had prior exposure but did not consult it during annotation. Requirement states were derived from AC labels (all ACs \textit{met} = Req. \textit{met}; all ACs \textit{unmet} = Req. \textit{unmet}; otherwise = Req. \textit{partially met}).

Before resolving disagreements, we measured inter-coder reliability across 150 requirements (5 apps × 30) for three raters (\textit{LLM-Specification}, \textit{Annotator 1}, \textit{Annotator 2}), indicating very high agreement (\textit{Krippendorff’s} $\alpha:$ 0.929, ordinal, 95\% CI [0.880, 0.966]). We adopt the agreed human evaluations as the gold standard; eight disagreements (out of 150) occurred between the LLM and humans.
At the acceptance-criteria level (5 apps x 30 requirements x 3 ACs), agreement between the two human annotators was similarly high (\textit{Cohen’s} $\kappa:$ 0.921, pooled, nominal, with accuracy: 96.7\%, 435 out of 450 ACs). Stratified by app, $\kappa$ ranged from 0.841 (\textit{App 1}) to 0.972 (\textit{App 2}), indicating high overall agreement with some heterogeneity.

\addvspace{0.3cm}
\noindent
\textit{Evaluation Setup and Results.} Based on our ground truth for each requirement and its acceptance criteria, we evaluated how well \textit{GUISpector} recognizes fulfillment and violations. For this, each app was analyzed in single evaluation runs. Requirements were classified into three categories (\textit{met}, \textit{partially met}, \textit{unmet}), while acceptance criteria were assigned two categories (\textit{met}, \textit{unmet}). We then calculated \textit{precision}, \textit{recall}, and \textit{F1} scores for the three-class requirements task and the binary acceptance-criteria task.

Table \ref{tab:results_nlr_pm_steps} summarizes per-app and average results. For the acceptance criteria task, both labels met (\textit{F1} avg. = 0.94) and unmet (\textit{F1} avg. = 0.87) achieve high scores, indicating that the framework reliably identifies fulfillment and violations at the criterion level.
For the three-class requirements task, \textit{F1} remains high for \textit{met} (avg. = 0.905) and \textit{unmet} (avg. = 0.878), whereas \textit{partially met} is slightly lower (avg. = 0.631), which may result from boundary confusions due to greater semantic ambiguity.
Execution effort varies by app (averg. \textit{steps} 19.5 for \textit{App 3} to 34.2 for \textit{App 5}). The observed spread is expected because each run starts from a clean state and certain apps require longer prerequisite flows before a requirement can be verified (e.g., always configuring a cleaning appointment prior to editing it). The elapsed \textit{time} per requirement likewise differs across apps for the same reasons. Although per-requirement runtime may, in some instances, exceeds that of human testers, the framework can make up for it by supporting straightforward parallelization.
Costs are dominated by MLLM input tokens and average \$0.67 per requirement, we expect reductions by merging verification runs.
\section{Related Work}

Automated GUI testing has evolved from early approaches such as automated test scripts \citep{xie2007designing}, random exploration \citep{mao2016sapienz}, and model-based techniques \citep{gu2019practical, zeng2016automated}, to more recent scripted bug replay methods \citep{gomez2013reran, feng2022gifdroid}. While these methods have improved testing efficiency, they often suffer from limited coverage due to the inherent complexity and dynamic behavior of GUIs. More recently, large language models (LLMs) have been leveraged to generate more human-like and semantically meaningful GUI interactions. For example, \textit{GPTDroid} \citep{liu2023chatting} frames GUI testing as a Q\&A task, using LLMs to iteratively explore app states, while \textit{CAT} \citep{feng2024enabling} combines retrieval-augmented generation and machine learning to automate UI test generation, and multi-agent approaches have been proposed for testing multi-user features \citep{feng2025agent}. However, these LLM-based methods typically rely on textual representations of GUI hierarchies, which neglect the visual and dynamic aspects of GUIs, limiting their ability to verify non-functional requirements and fine-grained acceptance criteria. In parallel, LLM-assisted prototyping tools \citep{kolthoff2024interlinking, kolthoff2025guide, kretzer2025closing} have introduced requirements matching and traceability features, but focus primarily on static, mid-fidelity GUI prototypes rather than fully interactive applications. Unlike these prior works, \textit{GUISpector} enables multi-modal, requirements-driven verification directly on interactive GUIs and uniquely introduces a closed feedback loop by providing actionable summaries from agent-based verification to inform and iteratively improve LLM-driven GUI implementation.
\section{Conclusion \& Future Work}

\textit{GUISpector} introduces a multi-modal LLM agent framework for automated verification of NL requirements in GUIs, enabling both interactive and fully automated workflows. Our approach demonstrates effective detection of requirement satisfaction and violations, and provides actionable feedback for iterative development. As future work, we aim to improve efficiency by merging related verification runs to avoid redundant exploration of shared subpaths.

\bibliographystyle{ACM-Reference-Format}
\bibliography{bibtex/bib/ref}

\end{document}